\documentstyle[fleqn,11pt]{article}

\def\mum{$\mu_r/M$}
\def\pj{\hspace{-.26cm}}
\def\fpj{\hspace{-.7cm}}
\def\thalf{{\textstyle{\frac{1}{2}}}}
\def\oneth{{\textstyle{\frac{1}{3}}}}
\def\tquar{{\textstyle{\frac{1}{4}}}}
\def\fiveth{{\textstyle{\frac{5}{3}}}}
\begin{document}
\title{Kaon Condensation in Neutron Star Matter with Hyperons}
\author{Paul J. Ellis\\
\small School of Physics and Astronomy\\[-3mm]\small University of
Minnesota\\[-3mm]
\small Minneapolis, MN 55455\\[2mm]
Roland Knorren and  Madappa Prakash\\ \small Physics
Department\\[-3mm]
\small State University of New York at Stony Brook\\[-3mm]\small Stony
Brook, NY 11794}
\date{~}
\maketitle
\begin{center}
\phantom{111}\vskip-2cm
{\small{\bf Abstract}}
\end{center}

Based on the Kaplan-Nelson Lagrangian, we investigate  kaon condensation in
dense neutron star matter allowing for the explicit presence of hyperons.
Using various models we find that the condensate  threshold is sensitive to the
behavior of the scalar density;  the more rapidly it
increases with baryon density, the lower is the threshold for condensation.
The presence of hyperons, particularly the $\Sigma^-$, shifts the threshold
for $K^-$ condensation to a higher density. In the mean field approach, with
hyperons, the condensate amplitude grows sufficiently rapidly that the nucleon
effective mass vanishes at a finite density and a satisfactory treatment of the
thermodynamics cannot be achieved. Thus, calculations of kaon-baryon
interactions beyond the mean  field level appear to be necessary.

\thispagestyle{empty}
{\small ~}
\newpage

The idea that, above some critical density, the ground state of baryonic matter
might contain a Bose-Einstein condensate of negatively charged kaons is due  to
Kaplan and  Nelson \cite{kapnel}. Subsequently, the formulation, in terms  of
chiral perturbation theory, was discussed  \cite{pol} and  the astrophysical
consequences have been explored in ref. \cite{tpl,refs}. Physically, the strong
attraction between $K^-$ mesons and   nucleons increases with density and
lowers the energy of the zero-momentum state. A condensate forms when this
energy becomes  equal to the kaon chemical potential, $\mu$.  In catalyzed
(neutrino-free)  dense neutron star matter,  $\mu$ is related to the electron
and nucleon chemical potentials by $\mu=\mu_e= \mu_n-\mu_p$ due to chemical
equilibrium  in the  reactions $n\leftrightarrow p+e^-+\bar{\nu}_e$  and
$n\leftrightarrow p+K^-$. The density at which this takes place is model and
parameter dependent, but is typically  $\sim4n_0$, where $n_0$ denotes
equilibrium nuclear matter density. Since  this may be less than the central
density in neutron stars, a $K^-$ condensate is expected to be present in the
core region. Apart from the softening effect on the equation of state, which
lowers the maximum  mass, the proton abundance is dramatically increased
resulting in a nucleon, rather than a neutron, star.

However, many calculations, e.g. \cite{glen,us}, of dense matter indicate that
hyperons, starting  with the $\Sigma^-$ and $\Lambda$, begin to appear at
densities $\sim(2-3)n_0$, a little lower than the kaon threshold mentioned
above. Once a significant number of negatively charged hyperons are present,
less electrons are required for charge neutrality, so the electron chemical
potential, $\mu_e$, begins to decrease with density.  Since $\mu_e=\mu$ this
would be expected to inhibit kaon  condensation, delaying it to a higher
density.    However, since overall charge neutrality must be maintained in  the
presence of many strongly interacting particles, the issue of  whether or not
kaon condensation would occur in matter containing hyperons is uncertain.   The
only previous work \cite{muto} involving both hyperons and kaons was concerned
with $p$-wave interactions which led to quasinucleons through the mixing with
hyperons.

Our objective here is to explore the influence of the explicit presence of
hyperons on kaon condensation in dense neutron star matter.  In our model
calculations, we employ the Kaplan-Nelson Lagrangian for the kaon-baryon
interactions and  a relativistic field theoretical approach  for the
baryon-baryon interactions.  Use of the relativistic approach for the baryons
allows for a distinction between the scalar and vector densities, which is
necessary for an adequate treatment of the kaon-baryon interactions.
Ignoring this distinction results in a lower threshold density for
kaon condensation.

We begin with the Kaplan-Nelson $SU(3)\times SU(3)$ chiral Lagrangian for
the kaons and the $s$-wave kaon-baryon interactions. Specifically
\begin{eqnarray}
{\cal L}_K&\pj=&\pj\tquar f^2 {\rm Tr}\partial_{\mu}U\partial^{\mu}U\!+
C{\rm Tr}\,m_q(U+U^{\dagger}\!-2)+i{\rm Tr}\bar{B}\gamma^{\mu}[V_{\mu},B]
+a_1{\rm Tr}\bar{B}(\xi m_q\xi+\!h.c.)B\nonumber\\
&&\pj+a_2{\rm Tr}\bar{B}B(\xi m_q\xi+h.c.)+a_3\left\{{\rm Tr}\bar{B}B\right\}
\left\{{\rm Tr}(m_qU+h.c.)\right\}\;.\label{pje0}
\end{eqnarray}
Here $U$ is the non-linear field involving the pseudoscalar meson octet from
which we retain only the $K^{\pm}$ contributions, $\xi^2=U$ and $B$ represents
the baryon octet -- nucleons plus hyperons. The quark mass matrix
$m_q={\rm diag}(0,0,m_s)$. For the mesonic vector current, $V_{\mu}$,
only the time component  survives in an infinite system with
$V_0=\thalf(\xi^{\dagger}\partial_0\xi+\xi\partial_0\xi^{\dagger})$. The
pion decay constant $f=93$ MeV and $C,a_1,a_2$ and $a_3$ are constants.
After some algebra, the relevant part of ${\cal L}_K$ takes the form
\begin{eqnarray}
{\cal L}_K&\pj=&\pj\left(\frac{\sin\chi}{\chi}\right)^{\!2}\Biggl\{
\partial_{\mu}K^+\partial^{\mu}K^-
+\frac{i}{4f^2}\frac{(K^+\partial_0K^--K^-\partial_0K^+)}{\cos^2\thalf\chi}
\sum_B(Y_B+q_B)B^{\dagger}B
\nonumber\\
&&\fpj
-\biggl(m_K^2+\frac{m_s}{2f^2}\sum_B\Bigl[(a_1+a_2)(1+Y_Bq_B)+(a_1-a_2)(q_B-Y_B)
+4a_3\Bigr]\bar{B}B\nonumber\\
&&\fpj+\frac{m_s}{6f^2}(a_1+a_2)(2\bar{\Lambda}\Lambda
+\sqrt{3}[\bar{\Sigma}^0\Lambda+\bar{\Lambda}\Sigma^0])
\biggr)\frac{K^+K^-}{\cos^2\thalf\chi}\biggr\}\;,\label{pje1}
\end{eqnarray}
where $\chi^2=2K^+K^-/f^2$, $q_B$ and $Y_B$ are the baryon charge and
hypercharge, respectively, and the kaon mass is given by $m_K^2=2Cm_s/f^2$.
We have not included in Eq. (\ref{pje1}) terms which simply give a constant
shift to the baryon masses; they indicate that $a_1m_s=-67$ MeV and
$a_2m_s=134$ MeV using the hyperon--nucleon mass differences. The remaining
constant $a_3m_s$ is not accurately known and we shall use values in the
range $-134$ to $-310$ MeV corresponding to 0 to 20\% strangeness content
for the proton. Some guidance is provided by recent lattice gauge simulations
\cite{liu} which find that the strange quark condensate in the nucleon is
large,
i.e. $\langle N|\overline ss|N\rangle = 1.16\pm0.54$. From the relation
$m_s\langle\overline ss\rangle_p = - 2(a_2+a_3)m_s$ and using $m_s=150$ MeV,
we obtain $a_3m_s = -(220 \pm40)$ MeV, which is in the middle of our range
of values.

In the baryon sector, we employ a  relativistic field theory model in which
baryons interact via the exchange of $\sigma$-, $\rho$- and
$\omega$-mesons. In the case that only nucleons are considered, $B=n,p$,
this is the well-known  Walecka model~\cite{sew} and we allow for
possible ``non-linear" $\sigma^3$ and $\sigma^4$ terms.  The Lagrangian
takes the form
\begin{eqnarray}
{\cal L}_B&\pj=&\pj\!\sum_B\bar B\left(i\gamma^{\mu}\partial_{\mu}-g_{\omega B}
\gamma^{\mu}\omega_{\mu}-g_{\rho B}\gamma^{\mu}{\bf b}_{\mu}\cdot
{\bf t} -\!M_B+g_{\sigma B}\sigma\right)\!B
-\tquar F_{\mu\nu}F^{\mu\nu}\!+\thalf m^2_{\omega}\omega_{\mu}\omega^{\mu}
\nonumber\\
&&\fpj
-\tquar {\bf B}_{\mu\nu}\cdot{\bf B}^{\mu\nu}+\thalf m^2_{\rho}{\bf b}_{\mu}
\cdot{\bf b}^{\mu}
+\thalf\partial_{\mu}\sigma\partial^{\mu}\sigma
-\thalf m^2_{\sigma}\sigma^2 - \oneth bM(g_{\sigma n}\sigma)^3
-\tquar c(g_{\sigma n}\sigma)^4\!.\label{hyp1}
\end{eqnarray}
Here $M_B$ is the vacuum baryon mass, the $\rho$-meson field is
denoted by ${\bf b}_{\mu}$,
the quantity ${\bf t}$ denotes the isospin operator which acts on the baryons
and the field strength tensors  for the vector mesons are given by the usual
expressions:--
$F_{\mu\nu}=\partial_{\mu}\omega_{\nu}-\partial_{\nu}\omega_{\mu}$,
${\bf B}_{\mu\nu}=\partial_{\mu}{\bf b}_{\nu}-\partial_{\nu}{\bf b}_{\mu}$.
The nucleon mass, $M=939$ MeV, is included in the penultimate term so
that $b$ is dimensionless.

The total hadron Lagrangian is then ${\cal L}_{\rm tot}={\cal L}_K+{\cal L}_B$.
We shall treat the kaons in the mean field approximation.
For the baryons, we shall consider calculations at the mean field level
(with the ``non-linear" terms so as to obtain a reasonable compression modulus
for nuclear matter) and at the one-loop Hartree level (without the
``non-linear" terms). As we shall see, there is a qualitative difference
between the results. We need to calculate the potential, $\Omega$, of the
grand canonical ensemble at zero temperature. Notice first that the
$\Lambda-\Sigma^0$ mass matrix needs to be diagonalized (for details see
ref. \cite{us}) producing eigenstates
\begin{equation}
H_1=\frac{\Sigma^0-\delta\Lambda}{(1+\delta^2)^{\frac{1}{2}}}\quad,\quad
H_2=\frac{\Lambda+\delta\Sigma^0}{(1+\delta^2)^{\frac{1}{2}}}\;,\label{mix}
\end{equation}
with corresponding masses $M^*_{H_1}$ and $M^*_{H_2}$. These are henceforth
included in the sum over baryon states $B$, along with the
$n,\,p,\,\Sigma^{+,-},\,\Xi^{0,-}$. The kaon-baryon interactions can be
included in the baryon effective masses and chemical potentials. The remaining
kaon terms are then easily treated  by writing \cite{tpl} the time dependence
of the fields $K^{\pm}=\frac{1}{\sqrt{2}}f\theta e^{\pm i\mu t}$; thus,
$\theta$ gives the condensate amplitude. Then it is
straightforward to obtain
\begin{eqnarray}
\frac{\Omega}{V}&\pj=&\pj
f^2(2m_K^2\sin^2\!\thalf\theta-\thalf\mu^2\sin^2\!\theta)+
\thalf m^2_{\sigma}\sigma^2+\oneth bM(g_{\sigma n}\sigma)^3
+\tquar c(g_{\sigma n}\sigma)^4\nonumber\\
&&\fpj-\thalf m_{\omega}^2\omega_0^2-\thalf m_{\rho}^2b_0^2
+\sum_B\left[\frac{1}{\pi^2}\int\limits_0^{k_{FB}}dk\,k^2(E_B^*-\nu_B)
+\Delta E(M^*_B)\right]\;.
\label{hyp2}
\end{eqnarray}
Here $V$ is the volume, $E_B^*=\sqrt{k^2+M^{*2}_B}$ and the effective masses,
$M^*_B$ for $H_1$ and
$H_2$ are obtained by diagonalizing the mass matrix, as we have mentioned,
while the remaining cases are given by
\begin{equation}
M^*_B=M_B-g_{\sigma B}\sigma+[(a_1+a_2)(1+Y_Bq_B)+(a_1-a_2)(q_B-Y_B)
+4a_3]m_s\sin^2\!\thalf\theta\;.
\label{emasses}
\end{equation}
The chemical potentials $\mu_B$ are given in terms of the effective chemical
potentials, $\nu_B$, by
\begin{equation}
\mu_B=\nu_B+g_{\omega B}\omega_0+g_{\rho B}t_{3B}b_0
-(Y_B+q_B)\mu\sin^2\!\thalf\theta\;,\label{hyp3}
\end{equation}
where $t_{3B}$ is the $z$-component of the isospin of the baryon and the
relation to the Fermi momentum $k_{FB}$ is provided by
$\nu_B=\sqrt{k_{FB}^2+M_B^{*2}}$.

In the one loop Hartree approximation (with $b=c=0$) , an additional
contribution  $\Delta E(M^*_B)$ to  the energy density is introduced due to the
shift in the single-particle energies caused by the negative energy baryon
states. After removing divergences,
$\Delta E(M^*_B)$ can be written~\cite{hr} in the form
\begin{eqnarray}
\Delta E(M^*_B)&\pj=&\pj-\frac{1}{8\pi^2}\biggl[4\left(1-\frac{\mu_r}{M}+
\ln\frac{\mu_r}{M}\right)M_B(M_B-M^*_B)^3-\ln\frac{\mu_r}{M}\;(M_B-M^*_B)^4
\nonumber\\
&&\qquad+M^{*4}_B
\ln\frac{M^*_B}{M_B}+M^3_B(M_B-M^*_B)-{\textstyle\frac{7}{2}}M^2_B(M_B-M^*_B)^2
\nonumber\\
&&\qquad\qquad\qquad+{\textstyle \frac{13}{3}}
M_B(M_B-M^*_B)^3-{\textstyle \frac{25}{12}}(M_B-M^*_B)^4\biggr]\;.
\label{walvac}
\end{eqnarray}
Here the necessary renormalization introduces a scale parameter, $\mu_r$. For
the standard choice~\cite{sew} of \mum=1, the first two terms in
Eq.~(\ref{walvac}) vanish.  Other choices of \mum  ~~introduce explicit
$\sigma^3$ and $\sigma^4$ contributions.  The
freedom to vary \mum\ allows the density dependence of the energy to be
modified, which makes it  possible~\cite{hr} to explore equations of state with
different stiffnesses.  We refer to this approach as the  modified
relativistic
Hartree  approximation (MRHA). In previous
work~\cite{pehr}, without hyperons, we found that while neutron star masses do
not significantly constrain  \mum, finite nuclei favor a value of 0.79.

The thermodynamic quantities can be obtained from the grand potential in Eq.
(\ref{hyp2}) in the standard way, thus the baryon number density
$n_B=(3\pi^2)^{-1}k_{FB}^3$, while for kaons
\begin{equation}
n_K=f^2(\mu\sin^2\!\theta+4e\sin^2\!\thalf\theta)\quad{\rm with}\quad
e=\sum_B(Y_B+q_B)n_B/(4f^2)\;.
\end{equation}
The pressure $P=-\Omega/V$ and the energy
density $\varepsilon=-P+\sum_B\mu_Bn_B+\mu n_K$.
The meson fields are obtained by extremizing $\Omega$, giving
\begin{eqnarray}
m_{\omega}^2\omega_0&\pj=&\pj\sum_Bg_{\omega B} n_B\quad,\quad
m_{\rho}^2b_0=\sum_Bg_{\rho B}t_{3B}n_B\;\quad {\rm and} \nonumber\\
m_{\sigma}^2\sigma&\pj=&\pj -bMg_{\sigma N}^3\sigma^2
-cg_{\sigma N}^4\sigma^3+\sum_B g_{\sigma B}n^s_B\;.
\label{hyp5}
\end{eqnarray}
Here $n^s_B$ denotes the baryon scalar density
\begin{equation}
n^s_B=\frac{1}{\pi^2}\int\limits_0^{k_{FB}} dk\,k^2\frac{M^*_B}{E_B^*}
+\frac{\partial\Delta E(M^*_B)}{\partial M_B^*}\;.
\end{equation}
The condensate amplitude, $\theta$, is also found by extremizing $\Omega$.
This yields the solutions $\theta=0$ (no condensate), or, if a condensate
exists, the equation
\begin{eqnarray}
&&\fpj\mu^2\cos\theta+2e\mu-m_K^2-d_1-d_2=0\qquad{\rm where}\;,\nonumber\\
&&\fpj2f^2d_1=\sum_{B\ne H_1,H_2}\Bigl[(a_1+a_2)(1+Y_Bq_B)
+(a_1-a_2)(q_B-Y_B)+4a_3\Bigr]m_sn^s_B\;,\nonumber\\
&&\fpj2f^2 d_2 \sin^2\!\thalf\theta =\sum_{B=H_1,H_2}(M^*_B
+g_{\sigma\Lambda}\sigma)n^s_B-M_{\Lambda}n^s_{H_1}
-M_{\Sigma}n^s_{H_2}\nonumber\\
&&\hspace{6cm}+\frac{1}{1+\delta^2}
(M_{\Lambda}-M_{\Sigma})(n^s_{H_1}-n^s_{H_2})\;,
\label{thresh1}
\end{eqnarray}
where we have taken $g_{\sigma\Lambda}=g_{\sigma\Sigma}$.

Finally we consider the leptons.  The electrons and muons are governed by
chemical potentials ~$\mu_e=\mu_{\mu}=\mu$.    (In the cold catalyzed state,
the antineutrinos would have left the star.)  Their contributions to the total
energy density and pressure are given adequately by the free gas expressions.
The requirement of chemical equilibrium in the weak processes yields
\begin{equation}
\mu_{H_1} = \mu_{H_2} = \mu_{\Xi^0} = \mu_n\ \ ;\ \
\mu_{\Sigma^-} = \mu_{\Xi^-} = \mu_n+\mu_e\ \ ;\ \
\mu_p = \mu_{\Sigma^+} = \mu_n - \mu_e \;. \label{murel}
\end{equation}
The remaining condition is that of overall charge neutrality, namely
\begin{equation}
\sum_B q_Bn_B-n_K-n_e-n_{\mu}=0\;.
\end{equation}

The couplings $g_{\sigma n}$ and $g_{\omega n}$ are determined by fitting the
equilibrium empirical properties of nuclear matter, while the coupling $g_{\rho
n}$ is obtained by fitting the nuclear symmetry energy.  For the mean field
model, we use the last two sets of values in table 2  of  ref.
\cite{glenmos}.  These models, termed models A and B hereafter, have
compression moduli of 300 and 240 MeV, respectively, and therefore exhibit
different stiffnesses for the high density equation of state. The couplings
employed for the MRHA model are shown in table 1.

Large uncertainties exist in the values of the hyperon--meson coupling
constants. We reduce the number of parameters by making the reasonable
assumption  that all the hyperon--meson coupling constants are the
same as those of the $\Lambda$.
Defining $x_\sigma=g_{\sigma\Lambda/}g_{\sigma n}$, with analogous definitions
for the $\omega$-- and $\rho$--couplings, the binding energy of the lowest
$\Lambda$ level in nuclear matter at saturation yields~\cite{glenmos}
\begin{equation}
-28=x_\omega g_{\omega n}\omega_0-x_\sigma g_{\sigma n}\sigma\;,\label{hyp6}
\end{equation}
in units of MeV. In ref. \cite{glenmos}, on the basis of fits to hypernuclear
levels and neutron star properties, the value $x_\sigma=0.6$ was suggested,
$x_{\omega}$ is then determined and the choice $x_\rho= x_{\sigma}$ was made.
We adopt this procedure, since we have examined
the effect of varying $x_\sigma$ and setting  $x_\rho=x_\omega$ and
found similar qualitative behavior.

To highlight the influence of hyperons on  kaon condensation, it is useful to
first consider the case where they are absent, i.e. nucleons-only matter.
Results for this case are
contained in table 2, and, in figures 1 and 2. The threshold density $n_c$
for condensation is determined from
\begin{eqnarray}
\mu^2 + \frac {(2n_p+n_n)}{2f^2} \mu - m_K^2
- \left [2a_1 n_p^s + (2a_2+4a_3) (n_p^s+n_n^s) \right] \frac {m_s}{2f^2}= 0\;,
\label{thresh2}
\end{eqnarray}
which is obtained by setting $\theta=0$ in Eq.~(\ref{thresh1}) and restricting
the sum over $B$ to $n$ and $p$ only.  Expressing $n_c$ as a ratio to
equilibrium nuclear matter density, $n_0$, we show in table 2 results for
$u_c=n_c/n_0$. The larger the magnitude of $a_3m_s$, the lower is the value
of $u_c$.  In addition, $u_c$ is systematically lower
for the MFT models than for the  MRHA models.  This latter feature is a
consequence of the more rapid  increase of  the scalar densities in the mean
field model than in the MRHA model (contrast panel (c) in fig. 1
and fig. 2).
Due to the rapidly growing condensate amplitude (see panel (e)),  the net
negative charge carried by the kaon fields grows rapidly with density
so that the leptons play only a minor role
in maintaining charge neutrality (the partial fractions are shown in  panel
(a)).  At high density, matter contains nearly equal abundances
of neutrons and protons. The softening induced by the presence of kaons has the
consequence that the maximum masses of stars are reduced from the case without
kaon condensation \cite{tpl,refs}.  It is worthwhile to point out that if the
distinction between the scalar and vector densities is ignored in
Eq.~(\ref{thresh2}),
condensation sets in at somewhat lower densities than those shown in table 2.

To summarize the case of nucleons-only matter: the onset of kaon condensation
is controlled by two factors, the first being the behavior
of the scalar density in dense matter and the second being
the magnitude of $a_3m_s$ which
is related to the strangeness content of the nucleon.
Additional work is needed to further pin down these quantities.

We turn now to the role of hyperons in matter, results for which are shown
in fig. 3 for the mean field model B with $a_3m_s=-222$ MeV.  Panel (a) of this
figure shows the composition.  One expects that
the $\Lambda$ and the $\Sigma^-$ first appear  at roughly the same density,
because the somewhat higher mass of the $\Sigma^-$ is compensated by the
presence of the electron chemical potential in the equilibrium condition of the
$\Sigma^-$ (see Eq.~(\ref{murel})).  More massive and more positively charged
particles than these appear at higher densities.  Use of Eq.~(\ref{hyp6}) to
constrain the hyperon couplings has the consequence that  the first strange
particle to appear is the  negatively charged $\Sigma^-$ hyperon.  The
$\Sigma^-$ competes with the leptons in  maintaining charge neutrality, so the
lepton concentrations begin to fall.   This is reflected in the electron
chemical potential $\mu_e$ (see panel (b)) which saturates at about 200 MeV and
begins to decrease from this value with increasing density.    Since
$\mu_e=\mu$, this has the consequence that kaon condensation now takes  place,
i.e. Eq. (\ref{thresh1}) is satisfied, at a higher density than in
nucleons-only matter.  This effect is seen for all the cases listed in
table 2.    Some insight into the role of the scalar densities may be gained by
examining the threshold condition Eq.~(\ref{thresh1}) in the presence of the
$\Sigma^-$ and $\Lambda$ hyperons only:
\begin{eqnarray}
\mu^2 + \frac {(2n_p+n_n-n_{\Sigma^-})}{2f^2}  \mu - m_K^2
- \left [2a_1 n_p^s + (2a_2+4a_3) (n_p^s+n_n^s+n^s_{\Sigma^-}) \right.
\nonumber \\
\hspace{2cm} \left. + \left( \fiveth (a_1+a_2)+4a_3 \right) n^s_\Lambda
\right] \frac {m_s}{2f^2}= 0\;.
\label{thresh3}
\end{eqnarray}
The first two terms in this equation are smaller than in the nucleons-only
case and this has to be compensated by the last term, which requires a higher
density. The increase in the critical density is less in the MFT
models,  since the scalar densities of the various particles increase more
rapidly than in the MRHA models.  In fact, for the MRHA the central density of
a
star containing all the baryons, but no kaons, ($u_{\rm cent}$ in table 2) is
less  than the critical density $u_c$ in almost all cases, which implies that
kaons will not be present. This is also the case for the MFT when the magnitude
of $a_3m_s$ is the smallest; however for larger values, which we favor, a kaon
condensate will be present in the star.

Returning to fig. 3, which corresponds to the MFT with $a_3m_s=-222$ MeV,
we see that in the presence of hyperons, the condensate amplitude increases
rapidly with density (panel (d)); so rapidly in fact that large changes are
induced in the scalar densities (panel (c)) and the Dirac effective masses of
all the particles (panel (b)).  The kaon contribution to the Dirac effective
masses of the nucleons
\begin{eqnarray}
M^*_p &=& M-g_{\sigma n}\sigma +(2a_1+2a_2+4a_3)m_s\sin^2\!\thalf\theta \,,
\nonumber \\
M^*_n &=& M-g_{\sigma n}\sigma +(2a_2+4a_3)m_s\sin^2\!\thalf\theta \,,
\end{eqnarray}
is negative and this causes the
effective masses to vanish at a finite baryon density.   Thus,  within the mean
field scheme adopted here and without an explicit treatment of a hadron to
quark matter transition, a calculation of the thermodyamics to encompass the
full range of densities in a neutron star is precluded.  This indicates
the need to consider several improvements:  (i) inclusion of
additional terms in the effective Lagrangian, such as the $p$-wave
interactions, (ii) inclusion of loop corrections in kaon-baryon interactions,
in particular an exact evaluation of the zero-point energy with the non-linear
Kaplan-Nelson Lagrangian.

In conclusion, we have shown that in dense matter the presence of hyperons
leads to kaon condensation at higher  densities than would otherwise be the
case. The critical densities depend sensitively on the behavior of the
scalar densities or, equivalently, the baryon Dirac effective masses in dense
matter.  A more rapid variation in these quantities results in lower thresholds
for condensation.  A mean field treatment of the kaon-baryon interactions
produces rather large effects so that the nucleon effective masses  vanish at a
finite baryon density, indicating unphysical behavior.   Clearly further work
is
needed to fully understand the role of the hyperon degrees of freedom on kaon
condensation in dense stellar matter.  \\

We thank Gerry Brown for much encouragement and for valuable discussions.  We
are grateful to Norman Glendenning for much helpful correspondence.  This
work was supported in part by the  U. S. Department of Energy under grant
numbers DE-FG02-88ER40328 (PJE) and DE-FG02-88ER40388 (MP and RK).

\centerline{\Large{\bf Figure Captions}}

\noindent Fig. 1.  Results are for nucleons-only matter in the mean field model
B with $a_3m_s=-222$ MeV.  (a) Relative fractions $Y_i=n_i/(n_n+n_p)$.
(b) Nucleon
Dirac effective masses, the kaon chemical potential $\mu=\mu_e$ and the scalar
field $\sigma$. (c) Scalar densities with $n^s = n_n^s+n_p^s$. (d) Pressure
with and without kaon condensation. (e) Condensate amplitude $\theta$ in
degrees. \\

\noindent Fig. 2.  Results are for nucleons-only matter in the Hartree
MRHA model with $\mu/M = 0.79$ $a_3m_s=-222$ MeV.
(a) Relative fractions $Y_i=n_i/(n_n+n_p)$. (b) Nucleon Dirac effective masses,
the kaon chemical potential $\mu=\mu_e$ and the scalar $\sigma$ field.  (c)
Scalar densities with $n^s = n_n^s+n_p^s$. (d) Pressure with and without kaon
condensation. (e) Condensate amplitude $\theta$ in degrees. \\

\noindent Fig. 3.  Results are for matter with hyperons in the mean field model
B with $a_3m_s=-222$ MeV.   (a) Relative fractions $Y_i=n_i/(\sum_Bn_B)$.
(b) Baryon
Dirac effective masses, the kaon chemical potential $\mu=\mu_e$ and the scalar
field $\sigma$. (c) Baryon scalar densities. (d)  Condensate
amplitude $\theta$ in degrees.

\newpage
\centerline{Table 1}
\centerline{MRHA coupling constants}
\begin{center}
\begin{tabular}{c|cccc|ccccc}\hline \hline
&\multicolumn{4}{c|}{Without hyperons}&\multicolumn{5}{c}{With hyperons}\\
${\displaystyle \frac{\mu_r}{M}}$&$K_0$&$C^2_{\omega}$&
$C^2_{\sigma}$&$C^2_{\rho}$&$K_0$&$C^2_{\omega}$&
$C^2_{\sigma}$&$C^2_{\rho}$&$x_\omega$\\ \hline
0.79&354&180.6&317.5&73.5&177&118.7&258.1&84.8&0.66\\
1.00&461&137.7&215.0&81.6&455&133.1&210.3&82.4&0.66\\
1.25&264&\phantom{1}78.6&178.6&90.8&228&\phantom{1}64.9&174.0&92.6&0.68\\
\hline \hline
\end{tabular}
\end{center}
\begin{quote}
Here $C_i^2=(g_{iN}M/m_i)^2$ and $K_0$ (in MeV) is the compression modulus.
The parameters are fitted to a nuclear matter binding energy of 16 MeV and a
symmetry energy of 30 MeV at a density of 0.16 fm$^{-3}$.
For the hyperon case we take $x_{\sigma}=x_{\rho}=0.6$.
\end{quote}
\centerline{Table 2}
\centerline{Critical density ratio, $u_c=n_c/n_0$, for kaon condensation}
\begin{center}
\begin{tabular}{lcc|ccc|cccr}\hline \hline
&&
& \multicolumn{3}{c|}{Without hyperons}
& \multicolumn{4}{c}{With hyperons} \\
&& $a_3m_s$
&  $-134$ & $-222$& $ -310$ & $-134$& $-222$ & $-310$ & \\
&&&&$u_c$&&&$u_c$&&$u_{\rm cent}^{a)}$\\ \hline
MFT& {\rm A} && 4.14 & 3.14 & 2.49 & 8.97 & 4.20 & 2.74 & 6.72\\
& {\rm B} && 4.15 & 3.15 & 2.49 & 9.46 & 4.22 & 2.73 & 7.66 \\ \hline
& $\mu_r/M=0.79$ && 5.22 & 4.54 & 3.89 & 15.3 & 9.87 & 7.32 & 7.81\\
MRHA& $\mu_r/M=1.00$ && 5.12 & 4.33 & 3.61 & 13.9 & 8.64 & 6.05 & 5.64\\
& $\mu_r/M=1.25$ && 5.24 & 4.55 & 3.93 & 20.8 & 13.4 & 9.57 & 8.44 \\
\hline \hline
\end{tabular}
\end{center}
\begin{quote}
$^{a)}$ $u_{\rm cent}=n_{\rm cent}/n_0$ is the central density ratio of a
neutron star containing
hyperons and nucleons in the absence of kaons. It can be compared with the
threshold density $u_c$ for kaon condensation.
\end{quote}


\begin{thebibliography}{99}
\bibliographystyle{unsrt}
\newcommand{\btem}{\bibitem}
\btem{kapnel} D. B. Kaplan and A. E. Nelson,
Phys. Lett. {\bf B175} (1986) 57.
\btem{pol} H. D. Politzer and M. B. Wise,
Phys. Lett. {\bf B273} (1991) 156;
G. E. Brown, K. Kubodera, M. Rho and V. Thorsson,
Phys. Lett. {\bf B291} (1992) 355.
\btem{tpl} V. Thorsson, M. Prakash and J. M. Lattimer,
Nucl. Phys. {\bf A572} (1994) 693.
\btem{refs} G. E. Brown, C-H. Lee, M. Rho and V. Thorsson,
Nucl. Phys. {\bf A567} (1994) 937;
T. Maruyama, H. Fujii, T. Muto and T. Tatsumi,
Phys. Lett. {\bf B337} (1994) 19, and references therein.
\btem{glen} N. K. Glendenning,
Ap. J. {\bf293} (1985) 470; Nucl. Phys. {\bf A493} (1989) 521.
\btem{us} M. Prakash {\it et al.},
Stony Brook preprint, SUNY-NTG-94-32 (1994).
\btem{muto} T. Muto,
Prog. Theor. Phys. {\bf89} (1993) 415.
\btem{liu} S-J. Dong and K-F. Liu, University of Kentucky preprint UK/94-07,
Lattice 94, Nucl. Phys. B (Proc. Suppl.), To be published.
\btem{sew} B. D. Serot,
Rep. Prog. Phys. {\bf55} (1992) 1855.
\btem{hr} E. K. Heide and S. Rudaz,
Phys. Lett. {\bf B262} (1991) 375.
\btem{pehr} M. Prakash, P. J. Ellis, E. K. Heide and S. Rudaz,
Nucl. Phys. {\bf A575} (1994) 583.
\btem{glenmos} N. K. Glendenning and S.A. Moszkwoski,
Phys. Rev. Lett. {\bf67} (1991) 2414.

\end{thebibliography}
\end{document}